# Estimation of unsteady aerodynamics in the wake of a freely flying European starling

Hadar Ben-Gida[1], Adam Kirchhefer[2], Zachary J. Taylor[1], Wayne Bezner-Kerr[3], Christopher G. Guglielmo[3], Gregory A. Kopp[2] and Roi Gurka[4]

[1] *School of Mechanical Engineering, Tel-Aviv University, Tel-Aviv, 69978, Israel*

[2] *Boundary Layer Wind Tunnel Laboratory, Faculty of Engineering, University of Western Ontario, London, N6A 5B9, Canada*

[3] *Department of Biology, Advanced Facility for Avian Research, University of Western Ontario, London, N6A 5B9, Canada*

[4] *Department of Mechanical Engineering, Ben-Gurion University of the Negev, Beer-Sheva, 84105, Israel*

**Abstract**

Wing flapping is one of the most widespread propulsion methods found in nature; however, the current understanding of the aerodynamics in bird wakes is incomplete. The role of the unsteady motion in the flow and its contribution to the aerodynamics is still an open question. In the current study, the wake of a freely flying European starling has been investigated using long-duration high-speed Particle Image Velocimetry (PIV) in the near wake. Kinematic analysis of the wings and body of the bird has been performed using additional high-speed cameras that recorded the bird movement simultaneously with the PIV measurements. The wake evolution of four complete wingbeats has been characterized through reconstruction of the time-resolved data, and the aerodynamics in the wake have been analyzed in terms of the streamwise forces acting on the bird. The profile drag from classical aerodynamics was found to be positive during most of the wingbeat cycle, yet kinematic images show that the bird does not decelerate. It is shown that unsteady aerodynamics are necessary to satisfy the drag/thrust balance by approximating the unsteady drag term. These findings may shed light on the flight efficiency of birds by providing a partial answer to how they minimize drag during flapping flight.

## 1  Introduction

Flapping flight is one of the most complex yet widespread propulsion methods found in nature. Although aeronautical technology has advanced remarkably over the past century, flying animals still demonstrate higher efficiency. One of the key open questions is the role of unsteady fluid motion in the wake of flying animals, and its contribution to the forces acting during the downstroke and upstroke [1]. The unsteady flow over small-scale wings has gained significant attention recently, both in the study of bird and insect flight, as well as to develop advanced aerodynamic models for high-performance micro-aerial vehicles [2]). The goal of the current study is to examine both the steady and the unsteady aerodynamics in the wake of a freely flying bird with a particular focus on the propulsive forces.

Among the first few attempts to describe unsteady aerodynamics [3-5], Brown [5] distinguished several patterns of flapping flight, and described complex movements of the wings through multiple sets of illustrations for different type of birds. Recently, Brunton and Rowley [6] developed reduced-order models for the unsteady aerodynamic forces on a small wing in response to agile maneuvers and gusts based on the framework suggested by Wagner [3] and Theodorsen [4]. However, they did not manage to augment their model with non-linear stall and separation models, which are important for improving lift modeling (see for example: Henningsson et al. [7]).

According to quasi-steady-state aerodynamic theory, slow-flying vertebrates should not be able to generate enough lift to remain aloft [8]. Therefore, unsteady aerodynamic mechanisms to enhance lift production have been proposed. Muijres et al. [9] showed that unsteady aerodynamic mechanisms are used not only by insects but also by larger and heavier fliers. Hubel and Tropea [10] verified Muijres et al.'s [9] findings by showing that the unsteady effects are not negligible for a goose-sized flapping model. Thus far, the main purpose of investigating unsteady aerodynamic mechanisms has been to understand their ability to enhance lift generation. However, it remains relatively unknown how unsteady aerodynamics participate in the drag and thrust balance of flapping flight.

The complex unsteady features of flapping flight introduce challenges to any realistic aerodynamic analysis. One of the first attempts to incorporate realistic wake structure in an aerodynamic model of bird flight was by Rayner [11-13] who proposed that each wingbeat was only aerodynamically active during the downstroke. As part of the wake structure, starting and ending vortices were suggested to be produced at the beginning and at the end of the downstroke phase, respectively. These vortices are connected by a pair of trailing vortices shed from the wingtips [11]. Therefore, at a certain distance downstream, the wake was assumed to be composed of a series of vortex rings referred to as 'elliptical loops' [12] which were conceptually related to the bird's wingspan and the circulation dictated by the force requirements (lift, profile drag and parasitic drag) of the bird. However, Rayner's model did not match later experimental observations. Spedding [14] performed measurements of vortex circulation in a jackdaw (*Corvus monedula*) wake and indicated that approximately half of the required momentum for weight support was present for jackdaw flight. As a consequence, the 'wake momentum paradox' arose. It was concluded that the discrepancy could be a result of unidentified complexities in the wake structure; i.e., Rayner's model was too idealized. Spedding [15] performed experiments, with the same apparatus as for the jackdaw on kestrel (*Falco tinnunculus*) flight at moderate ($U_\infty$=7 m/s) speeds and observed a distinctly different wake topology. He suggested that instead of discrete loops separated by aerodynamically inactive upstrokes, two continuous undulating vortex tubes were found in the wake; i.e., the upstrokes were also aerodynamically active. The measured circulation of the shed vortices was similar during the downstroke and upstroke, and it was adequate for supporting the weight of the bird. In addition to the lift force, it was found that generating net thrust occurred through varying the wing geometry, not through varying the circulation, implying that the wing motion is important.

The concept of different wake topologies has prompted analyses of vortex gaits [16-18]. Rayner [16] characterized the vortex gait selection for flying birds and stated that the choice between the different vortex gaits is determined by flight speed and wing morphology. Spedding et al. [18] investigated the wake structure behind a thrush nightingale (*Luscinia luscinia*). They concluded that the structure of the wake downstream, far from the body (roughly 17 chord lengths), varies gradually from an approximately elliptical vortex wake to a continuous trailing vortex wake. Therefore, the wake is comprised neither of a series of elliptical vortex loops, nor a pair of continuous trailing vortices, but is a combination of both. Eventually, the 'wake momentum paradox' was addressed, within the bounds of experimental uncertainty, through the high spatial resolution available in PIV, as well as a detailed accounting procedure for the calculation of circulation in the wake [18]. However, the relationship between specific wake topology and the propulsive aerodynamics of bird flight remains an open question.

The vorticity structure in the wake of a flying bird, similar to the distribution of vorticity in any wake, is dependent on the boundary conditions. Consequently, the kinematics of the wings has been investigated in the literature with the goal of either using the wing motion to predict forces or associating the wake topology with the motion of the wings. Two wingbeat kinematics of a thrush nightingale [19], and of two individual robins [20], were quantified in order to relate them to their vortex wakes. However, the kinematic variations with flight speed occurred only during the upstroke period where the wing folding and the wingbeat frequency were observed to vary. In addition to the wingbeat kinematics, the streamwise distance between the bird and the measurement location has been shown to be of importance when drawing conclusions about the aerodynamics related to the wake structure.

Hedenström et al. [20] investigated wakes behind European robins (*Erithacus rubecula*) and found that they resemble the thrush nightingale wake [18]. It was argued that the wakes' circulations were similar because the measurements were in the far wake, significantly downstream of the bird [1]. Studying both the near and far wake of Pallas' long tongued bat (*Glossophaga soricina*), Johansson et al. [21] concluded that measurements in the far wake might lead to misinterpretation of the wake topology. This misinterpretation occurs because the near wake is more readily tied to the generating wing kinematics and, thus, contains details of vortex structures that could easily be missed in the far wake [21].

All the former work described here does not take into account the unsteady portion of the flow presumably generated by the wing's motion. Rayner et al. [22] performed measurements on starlings in undulating flight in a wind tunnel and showed that the geometry of the flight path depends upon wingbeat kinematics, and that neither the flapping nor the gliding phases of flight occur at constant speed or at constant angle to the horizontal. The bird gains both kinetic and potential energy during the flapping phases making it difficult to model. Rayner et al. [22] indicate that such speed variation can provide significant savings in mechanical power in both bounding and undulating flight. Recently, high-speed PIV systems have become available for animal flight research with which the wake structure can be analyzed at a high temporal resolution [23, 24]. In these studies, the wake is sampled using PIV images taken at a typical frequency of 200 Hz in a transverse plane (vertical spanwise) referred to as the Trefftz plane [25]. The three-dimensional wake is assembled by identifying coherent streamwise structures, such as the tip vortex [26], and, in consequence, the time varying flight forces have been estimated based on this method [27-30]. However, the focus in such studies has been on the lift force and not on the relation between drag and thrust. It is also noteworthy that PIV measurements in the Trefftz plane consist of substantial uncertainty in the estimated velocity and its gradients due to the set-up complexity and the nature of the PIV technique when performing flow measurements with a strong out of plane velocity component [25]. Therefore, conclusions drawn based on these measurements should be carefully utilized. To date, there is no available volumetric technique capable of performing full three-dimensional measurements of high speed flows in air. The existing techniques such as StereoPIV provide three velocity components but not gradients. In addition, the accuracy level of this technique in reconstructing the third dimension is not high [31, 32]. Therefore, the current models and quantitative estimations of forces behind the wake in the Trefftz plane are subject to relatively large error [25].

Currently, most of the aerodynamic models for birds are based on fixed wings in steady flow [33]. While the quasi-steady values seem to be valid in many bird wakes, it is far from proof that the approach is valid. Therefore, the current study addresses the near wake variations behind a freely flying bird in time and space with a particular focus on the unsteady aerodynamics of the flow that results from the flapping wing. The change of velocity with time is the key parameter that marks the unsteady effect and its variation is examined in terms of the drag and thrust balance.

## 2 Experimental setup

### 2.1 Wind tunnel

The experiments were performed in the closed-loop hypobaric climatic wind tunnel at the Advanced Facility for Avian Research (AFAR) at the University of Western Ontario. The test section of this wind tunnel has a cross-sectional area of 1.2 m$^2$, is preceded by a 2.5:1 contraction, and is enclosed in a hypobaric chamber. The width, height and length of the test section are 1 m, 1.5 m, and 2 m, respectively. An open jet exists between the downstream end of the test section and the diffuser for the purpose of introducing the live bird into the wind tunnel during the experiments. The turbulence intensity is lower than 0.3% at the location where the measurements were taken. A fine net was placed at the upstream end of the test section to prevent the bird from entering the contraction, which was not observed to alter the turbulence significantly. The flight conditions were at atmospheric pressure, a temperature of 15 °C, and relative humidity of 80%.

### 2.2 The Bird - European Starling

The wake measurements (as illustrated in Figure 1) were taken from a European starling that had been trained to fly in the AFAR wind tunnel. The bird's wings had an average chord, $c$, of 6 cm, a maximum wingspan of $b$=38.2 cm and an aspect ratio (wingspan squared divided by the wings lifting area), $AR$, of 6.4. A typical cruising speed of $U_\infty$=12 m/s was chosen for the experiments based on the comfort of the starling and its ability to fly for prolonged periods of time during the testing. The wingbeat frequency, $f$, was 13.3 Hz on average, and the average peak-to-peak wingtip vertical amplitude, $A$, was 28 cm. These quantities correspond to a chord-based Reynolds number of $4.8 \cdot 10^4$, a Strouhal number, $St=Af/U_\infty$=0.30, and a reduced frequency, $k=\pi fc/U_\infty$=0.20. At the time the experiments were performed the bird had a mass of 78 g and a lateral body width of 4 cm.

Due to the powerful laser operating within a few chord lengths of the bird's tail, two precautions were taken to ensure the bird's safety. Goggles made of a flexible, optically dense, polymer material (Yamamoto Cogaku Co YL 600) were designed to protect the bird's vision as well as to reduce the potential of the light sheet frightening the bird. After an accommodation period of 20 minutes in a cage, the bird would fly normally in the wind tunnel while wearing the goggles. In addition, for preventing direct contact between the bird and the light sheet, a collection of optoisolators operated by six infrared transceivers were integrated into the PIV system. The function of the optoisolators was to trigger the laser only when the bird was in a desired position upstream of the PIV field of view, thus ensuring that the bird was in a position where it was not in danger of being hit by the laser. All animal care and procedures were approved by the University of Western Ontario Animal Use Sub-Committee (protocols 2006-011, 2010-216).

*2.3 Flow velocity and kinematic measurements*

Flow measurements were taken using the long-duration time-resolved PIV system developed by Taylor et al. [34]. Olive oil particles, 1μm in size [35] were introduced into the wind tunnel using a Laskin nozzle from the downstream end of the test section so that it did not cause a disturbance to the flow in the test section or to the bird. The PIV system consists of an 80 W double-head, diode-pumped, Q-switched, Nd:YLF laser at a wavelength of 527 nm and two CMOS cameras (Photron FASTCAM-1024PCI) with spatial resolution of 1024x1024 pixel$^2$ at a rate of 1000 Hz. The PIV system is capable of acquiring image pairs at 500 Hz using the two cameras for 20 minutes continuously. In the current experiments, one camera was used to record the bird kinematics during the wingbeat and the other was used for PIV measurements in the wake. The PIV camera's field of view was approximately 12 x 12 cm$^2$ in size, or 2*c* by 2*c*. Vector fields were computed by OpenPIV [34] using 32 x 32 pixel$^2$ interrogation windows with 50% overlap, giving a spatial resolution of 32 vectors per chord. In the current experiments, 4 600 vector maps were recorded, and out of this dataset, 650 vector maps contained features of the near wake behind the starling's wing. The PIV data were measured 4 wing chord lengths (~0.24 m) behind the right wing, and therefore it took 20 ms for events generated at the wing to enter the PIV field of view. The wake was sampled in the parasagittal plane (9 x 9 cm$^2$) at 2 ms intervals (500 Hz), so that both the downstroke and upstroke phases were temporally resolved.

The streamwise and vertical positions of the bird for all measurements were recorded simultaneously with the flow field measurements. The field of view in these recordings had an area of 9*c* by 9*c*. Figure 2 depicts a sample image of the starling flying in the tunnel as captured by the camera. The box marked with "PIV" indicates the location of the measured velocity fields from the PIV system. In addition, a floor-mounted camera operating at 60 Hz was used to record the spanwise position of the bird as well as the laser sheet illumination. These images allowed for the identification of the measured PIV plane in respect to the position of the wing; therefore, the wake velocity field could be associated with the spanwise location across the wing or the body. The floor-mounted camera was not synchronized with the PIV system; therefore, the two time histories were synchronized manually based on the presence of light from the laser firing in the images. Once synchronized, spanwise positions were assigned to the wake data captured at 500 Hz based on interpolation from the simultaneously recorded spanwise positions recorded at 60 Hz.

An error analysis based on the root sum of squares method has been applied to the velocity data and the wing kinematics. The errors were estimated as: 2.5% for the instantaneous velocity values, 12% for the instantaneous vorticity and 3% for the drag values [36]. The error introduced in the kinematic analysis resulted from the spatial resolution of the image and the lens distortion leading to an estimated error of 5% in the wing displacements.

## 3   Results

A number of wake velocity maps were sampled where the starling was flying in a steady flapping mode without performing any maneuvers. The data discussed herein was selected from a broad acquisition batch where the bird was flying continuously for a few minutes (see Figure 3, where the streamwise velocity at the wake is depicted as a time series). The selection criterion was based on the flight mode chosen: no net acceleration of the bird over a wingbeat cycle, as observed from the high speed imaging. The wing kinematics and the flow analysis are presented in the following sections. The analysis includes four sets of wingbeats each comprising a downstroke phase and an upstroke phase. The first three wingbeat sets contain a total of 110 velocity maps and kinematic images acquired simultaneously. These sets feature three consecutive wingbeats (referred to as wingbeats 1, 2 and 3, according to their order of appearance). A fourth wingbeat set (wingbeat 4) contains 43 vector maps and kinematic images.

*3.1   Kinematic Analysis*

Figure 4 illustrates the starling in different positions during wingbeats 1, 2, 3 and 4. Since the purpose of this study is to estimate the streamwise forces acting on the bird, it is imperative that the bird accelerates negligibly in this direction. From the kinematics shown in Figure 4 it is observed that, during all four wingbeats, the starling does not accelerate noticeably in the streamwise direction. Thus, analysis of the wake aerodynamics during these wingbeats can be performed assuming negligible acceleration of the bird in the streamwise direction.

In order to characterize the kinematics of the bird, several parameters have been estimated which are commonly used to evaluate bird flight characteristics beginning with the flapping frequency and Strouhal number [2] (see Table 1). The flapping frequency was calculated according to the inverse of the period, $T$, of each wingbeat $f = 1/T$, and the Strouhal number according to

$$St = \frac{fA}{U_\infty}. \tag{1}$$

where $A$ is the amplitude of the wingtip and $U_\infty$ is the free stream velocity. Taylor et al. [37] suggest that, for a wide variety of animals (including fish, birds and insects) efficient cruising locomotion requires that the Strouhal number be in the range of $0.2 < St < 0.4$, and for birds during cruising flight it should be nearly 0.2. The current results fall within this predicted range, and the minimum value ($St = 0.24$) approaches the value for cruising flight during wingbeat 4.

The wingtip angle of attack, $\varphi$, is computed by assuming that the wing moves up and down through its amplitude at a constant vertical speed

$$v = 2fA. \tag{2}$$

For a wing moving forwards with a streamwise velocity $U_\infty$, the wingtip will move either up or down with a maximum 'zigzag' angle given by [38]:

$$\tan(\varphi) = \frac{2fA}{U_\infty} \tag{3}$$

Substituting the Strouhal number from Eq. (1) yields:

$$\varphi = \tan^{-1}(2St). \tag{4}$$

For a bird flapping its wings up and down in the vertical plane and keeping the wing chord horizontal all the time (e.g., Figure 4), the wingtip angle of attack can be approximated as $\varphi$ on the downstroke and $-\varphi$ on the upstroke [38]. It is observed that the minimum mean wingtip angle of attack during the downstroke phase ($\varphi=26º$) occurs for wingbeat 4 and that the value is larger than the stalling angle of conventional fixed wing aircraft of approximately 15º [39]. However, as shown by Nachtigall and Wieser [40], the angle of attack varies from zero at the shoulder to a maximum value at the wingtip. Thus, on average, the attack angle over much of the wingspan is lower than the stalling angle of 15º.

*3.2    Wake characteristics*

In the previous section, the wing kinematics were quantified for each of the four complete wingbeats.  In this section, the wake is characterized in terms of aerodynamic forces and vorticity content. The vorticity in the wake is computed directly from the PIV data using a least squares differentiation scheme [36]. To determine if the vorticity as measured at the near wake is sufficient for the force estimations, the peak vorticity in the current data behind the starling is compared with former works [1, 7, 18, 20, 41], as depicted in Figure 5. The peak spanwise vorticity measured in the wakes of several flapping wing animals is displayed in Figure 5 for the purpose of contextualizing the current measurements in the starling wake among other flapping wing animal studies. In Figure 5, the spanwise vorticity is normalized by the mean chord and wind speed of each respective study. Since peak vorticity measured in the wake of a cruising animal varies gradually over the range of flight speeds [18], peak values of spanwise vorticity are included for both extremes of the natural speed range where possible.

It is observed that the peak normalized vorticity (4.1) as depicted in Figure 5 of the starling in the present study is larger than values of peak vorticity from animals in cruising or fast flight (red and purple bars). Peak values of vorticity from birds and bats flying at the lower end of their natural speed range (blue and green bars) are more comparable to what is displayed by the starling.

*3.2.1    Wake Reconstruction*

During flapping flight, bird wings change position causing the momentum and circulation in the wake to vary. In many simplified models, the wake changes in a periodic manner where the downstroke and upstroke phases have different signatures [12, 18, 43]. In order to characterize the effect of the flapping action on the near wake behind the starling and its impact on the aerodynamic performance, sequences of velocity maps have been reconstructed. This procedure was performed using PIV data collected at a sampling rate of 500 Hz – significantly higher than the 13.3 Hz wingbeat frequency. Therefore, a pattern of vorticity appearing in one frame also appears in the consecutive frame; only phase-shifted. The wake composite is formed by plotting sequential instantaneous vorticity fields computed from PIV data and by matching patterns in the vorticity fields with a shift. The offset of the $n^{th}$ successive PIV images is calculated as $U_c \cdot \Delta t \cdot n$. The convection velocity, $U_c$, is the velocity at which the characteristics of the wake collectively travel downstream. In the present study, wake composites have been generated using the free-stream velocity ($U_\infty$) as a convection velocity. The generation of a wake composite provides a useful visualization tool for observation of the wake dynamics over the time series of a wing beat cycle. What appears as "downstream" in the wake composite happens earlier, while what appears "upstream" in the composite happens later meaning that the generation of the wake composite invokes Taylor's hypothesis in which the characteristics of the flow are a frozen spatial pattern advected through the field of view.

Figure 6 demonstrates the wake reconstruction procedure: initially two consecutive spanwise vorticity fields, along with the velocity fluctuations, are put side by side. Afterwards, a vorticity pattern classification process is performed on each image, which eventually assists in the identification of similar patterns between the two images according to similarity in size, shape, direction and value. Figure 6 shows four different negatively-signed (*A-D*) patterns and two different positively-signed (*E* and *F*) patterns. It can be seen that the different patterns of vorticity move downstream during the time difference between the consecutive images (2 msec).

### *3.2.2  Wake Evolution*

The wake features are presented through fluctuating velocity and vorticity fields as depicted in Figure 7. The set of figures describes the wake of the freely flying starling during the four different wingbeats (as defined in §3.1). Each wake pattern consists of 24 consecutive fields displaying the spanwise vorticity, $\omega_z(x, y)$, varying from -650 to 650 $sec^{-1}$. In addition, the spatially averaged velocity has been subtracted from each frame, so the velocity vectors displayed are fluctuations.

Using the floor-mounted camera, the wake patterns presented in Figure 7 were determined to be captured in a plane that is, on average, approximately 2.5 cm from the right wing root (14% of the wing length). Figures 8 and 9 demonstrate the different wing sections being intersected with the laser sheet as a consequence of the starling's small spanwise movements throughout the wingbeats. It should be noted that, in the four wingbeats presented in this study, the starling was recorded with the minimum possible spanwise, vertical and streamwise movements.

The most immediate observation from the wake reconstructions in Figure 7 is the periodicity of the wake over the shedding cycle. In light of the topology of spanwise vorticity, it is convenient to discuss the wake in terms of a top half and a bottom half where the wake center is defined by the location of greatest velocity deficit. As would be expected, the top half of the wake is composed primarily of negative spanwise vorticity and the bottom half of the wake by positive spanwise vorticity. It is also noted that the quantity of lift producing vorticity (i.e., negative) is greater than that of positive vorticity.

In the available literature, there has been considerable focus on the vortex topology in bird wakes [18, 20, 28]. A comparison with these works shows that there are qualitative similarities in the vorticity structure between the current measurements close to the body (along the span) and those shown by Henningsson et al. [7] for a swift. From their measurements of the wake at 10 chord lengths downstream of a flying swift, Henningsson et al. [7] suggest that the tip vortices for the swift are connected by spanwise vortices. The measurement plane in the current study is significantly closer to the bird in the streamwise direction (4 chord lengths) offering new perspective on wake development. Distinguishing coherent vortices from shear in a real fluid flow is not trivial [42-44]. Many wakes are typified by high shear and bird wakes are no different. Vorticity alone has shown to be inadequate in distinguishing a vortex from an area of high shear [43], and the PIV data in the current study are of insufficient resolution to distinguish if the majority of the vorticity observed in Figure 7 is due to coherent vortices or high shear. Thus, if there are spanwise connecting vortices, they should be relatively small at the location of our measurement plane. The strong vorticity observed in the wake implies the presence of shear and, as a result, drag. In the next section, the sectional drag force is examined quantitatively.

## 3.3 Drag estimates

Consider a section of the bird wing as a two-dimensional body in an incompressible flow as sketched in Figure 10. The body is located within a control volume (*abcsdefghia*) with its width in the *z* direction being unity. Inside the control volume, the integral form of the momentum equation is [45]

$$\frac{\partial}{\partial t}\iiint_V \rho \vec{u} \, dV + \iint_S (\rho \vec{u} \cdot d\vec{S})\vec{u} = -\iint_{abhi} p \, d\vec{S} - R' \tag{5}$$

where $\vec{u}$ is the velocity, $\rho$ is the density (constant), $p$ is the pressure and $R'$ is the resultant aerodynamic force per unit span exerted on the body by the normal and shear stresses acting at the body surface. Note that the viscous terms have been ignored as they scale with Re$^{-1}$. The integrals are taken over the control volume, $V$, enclosed by the surface, $S$. Using the $x$-component of Eq. (5), the aerodynamic drag per unit span, $D'$, is

$$D' = -\rho \frac{\partial}{\partial t} \iiint_V u \, dV - \rho \iint_S (\vec{u} \cdot dS) u - \iint_{abhi} (p \, dS)_x \qquad (6)$$

A positive value of $D'$ is defined as drag and a negative one as thrust. For simplicity, the resultant force, $D'$, is referred to as drag per unit span where negative drag per unit span refers to thrust. For $S$, estimated sufficiently far from the body where the pressure is assumed constant, and equal to the undisturbed free-stream pressure $p_\infty$, Eq. (6) becomes:

$$D' = -\rho \frac{\partial}{\partial t} \iiint_V u \, dV - \rho \iint_S (\vec{u} \cdot dS) u \qquad (7)$$

By definition, $\vec{u}$ is parallel to the streamlines and $dS$ is perpendicular to the control surface. Thus, for streamlines $ab$, $hi$ and $def$, the multiplication $\vec{u} \cdot dS = 0$. In addition, the planes $cd$ and $fg$ are adjacent to each other, so their contribution to the second term in Eq. (7) cancel each other. As a result, the second term in Eq. (7) consists of contributions only from sections $ai$ and $bh$ (where $dS = dy$). Therefore, Eq. (7) is expressed as

$$D' = -\rho \frac{\partial}{\partial t} \iint_{S_{abhi}} u \, dx \, dy - \left( -\rho \int_i^a u_1^2 \, dy + \rho \int_h^b u_2^2 \, dy \right) \qquad (8)$$

Using the integral form of the continuity equation and multiplying by $u_1$ (a constant in the current case),

$$\rho \int_i^a u_1^2 \, dy = \rho \int_h^b u_2 u_1 \, dy \qquad (9)$$

and substituting Eq. (9) into Eq. (8) leads to

$$D' = -\rho \frac{\partial}{\partial t} \iint_{S_{abhi}} u \, dx \, dy + \rho \int_h^b u_2 (u_1 - u_2) \, dy \qquad (10)$$

Therefore, the drag is composed of two terms: steady (second term) and unsteady (first term). The steady drag per unit span, referred to as the velocity deficit drag in classical aerodynamics, can be derived from the second term in Eq. (10) and expressed as

$$D'_{Steady} = \rho \int_0^h u(U_\infty - u) \, dy \qquad (11)$$

where $h$ is the wake vertical extent of the PIV velocity field. The so-called 'unsteady drag' per unit span can be derived using the first term in Eq. (10) and expressed as

$$D'_{Unsteady} \approx -\rho \frac{\partial}{\partial t} \int_0^h \int_0^l u \, dx \, dy \qquad (12)$$

where $l$ is the streamwise extent of the PIV velocity field.

It should be noted that the full area integral (or, volume integral per unit span) cannot be directly computed as it appears in Eq. (10) since only a portion of the control surface enveloping the bird wing is measured in the current experiments (e.g., Figure 2). However, Eq. (12) is used as an approximation to the entire area integral bound by the control surface shown in Figure 10. Figures 11 and 12 describe the time variation of the steady and unsteady drag per unit span as computed from Eq. (11) and (12) for the four different wingbeats (see a-d in Figures 11 and 12) as depicted in Figure 7. The integrals were performed for each instantaneous velocity field of the starling's wake. For the steady drag per unit span, different streamwise velocity profiles were sampled at different $x$-positions for each velocity field map. Subsequently, these profiles within one PIV vector map were spatially averaged into one profile describing the velocity deficit (steady drag per unit span) similar to the procedure described in [18]. This procedure is similar to a spatial windowing average in order to smooth out some of the variations within each vector map, and it is noted that the general trend over the wingbeat cycles does not change using this procedure. Each point in Figures 11 and 12 represents the integral value depicted from each velocity field yielding a time evolution of the drag in the near wake. Figures 11e and 12e depict the averaged drag profiles for the four wingbeats. The uncertainty of the computed drag values in Figures 11 and 12, estimated in §2.3, is similar to the size of the markers. Notable differences are observed between the steady and unsteady components of the horizontal momentum, shown in Figures 11 and 12. Figure 11e shows that the averaged steady drag is positive over the entire wingbeat cycle except for a short period in the transition from downstroke to upstroke. Contrary to the steady portion, the unsteady contribution to the drag as depicted in Figure 12e is negative during both the downstroke and upstroke while positive during the transition phase. The steady drag values reach 1 and 0.5 N/m during the upstroke and downstroke, respectfully. The negative values as calculated for the unsteady portion reach a minimum value of -0.5 N/m. These differences are discussed in detail in the following section.

## 4  Discussion

In classical aerodynamics, the profile drag (Eq. (11)) inherently assumes that $\partial u/\partial t = 0$ everywhere in the chosen control volume. While this assumption is reasonable in the wake of a section model mounted in a wind tunnel, it seems unlikely that this condition is satisfied in the wake of a freely flying bird. However, experimental measurements of the entire control volume remain prohibitive and many studies have approximated the drag force through the profile drag [18, 46]. In the current study, the profile drag has been estimated in the same manner over four complete wingbeats of a freely flying starling (Figure 11). The steady drag was demonstrated to vary significantly during the wingbeat cycle yielding higher profile drag during the upstroke. This variation in the drag force over the wingbeat cycle is consistent with earlier studies [47], which suggested that the upstroke phase generates more drag than the downstroke phase.

The profile drag developed over the wing is manifested through the wake velocity profile. In the case of flapping wings (or in any event where the flow is disturbed by some external motion), the velocity field changes spatially and temporally. Since the bird is flying freely, and the kinematic images (Figure 4) demonstrate that the bird does not accelerate in the streamwise direction, there is no net momentum change in the streamwise direction. Since the profile drag is the term relevant to the overall streamwise momentum balance of Eq. (5) then one would expect a momentumless velocity profile (see Figure 13). However, as indicated by the profile drag, which is generally positive (Figure 11e), as well as in profiles presented in previous studies [18], it appears that another force is required to balance the streamwise momentum.

From the momentum balance (Eq. (5)) it is expected that the compensating force for the profile drag is the volume integral containing the $\partial u/\partial t$ term. Considering the control volume drawn in Figure 10, it is plausible to assume that $\partial u/\partial t = 0$ everywhere at the flow upstream of the wing since the wind tunnel is operating at a constant wind speed. Conversely, $\partial u/\partial t$ is not expected to be zero around the wings and especially in the wake. It is expected that $\partial u/\partial t$ is greater than zero in the wake during the downstroke since work (e.g., the flapping motion) is an input to the wake [45]. However, during the upstroke it remains unclear if this term is negative or positive. The estimate of the 'unsteady drag' (Eq. (12); Figure 12) uses a portion of the volume as an approximation to the total volume integral of $\partial u/\partial t$. The results presented in Figure 12e demonstrate that $\partial u/\partial t$ is smaller than zero as expected during the downstroke as the bird generates lift and propels itself forward [45]. At the beginning of the upstroke, the volume integral of $\partial u/\partial t$ appears to be positive; however, once the steady drag reach a maximum value halfway through the upstroke (Figure 11e), the unsteady drag once more becomes negative indicating that $\partial u/\partial t > 0$ as it was during the downstroke.

It should be noted that performing PIV measurements around freely flying birds limits the capability of capturing the entire span of the wing; therefore, our estimates are based on a sectional measurement. However, the current dataset offers much lower uncertainty of the streamwise velocity compared to attempts to capture the entire volume using the Trefftz plane [25]. Drag variations may also arise due to the wing flexing, which acts to minimize the drag during the second half of the upstroke phase [2]. In addition, it is well known that a significant amount of thrust is generated at the outer part of the wing [2]. However, the time-resolved streamwise velocity measurements in the near wake of a freely flying starling suggest that for a complete understanding of the drag and thrust relationship in bird wakes the importance of $\partial u/\partial t$ cannot be neglected. Furthermore, the flow mechanisms underlying the present measurements of $\partial u/\partial t$ are expected to correlate to the bird's use of unsteady aerodynamics and improved efficiency.

Many studies that investigate flying animals utilize quasi-steady models to analyze the flapping mechanism [16, 18, 45]. Quasi-steady analysis of flapping flight generally assumes that the aerodynamic forces in flapping flight can be composed from the various instantaneous wing configurations, as they would behave in an equivalent series of steady flows. However, our estimation of the steady drag force shows that it is mostly positive indicating that the bird should be decelerating if acted on by this force alone. The kinematic observations show that the bird is not accelerating in the streamwise direction, so there should be a balancing force that is not accounted for in the steady aerodynamics. The approximation of the unsteady contribution to the streamwise force indicates that the balancing force to the steady velocity deficit drag is most likely due to unsteady aerodynamics, thus revealing an inadequacy of quasi-steady approaches. Future studies measuring a greater spanwise distribution of drag forces are necessary to obtain the complete description of steady versus unsteady aerodynamics.

## 5 Conclusions

In this study, a long-duration time-resolved PIV system was used to obtain accurate, time-resolved measurements of the streamwise velocity in the wake of a freely flying European Starling flying in flapping flight at the AFAR hypobaric wind tunnel. The system is capable of capturing images for 20 min continuously in order to characterize unsteady phenomena within a given flow field. A total of 4,600 vector maps were analyzed in the current study, and four wingbeat cycles were identified within this data set. The identification was performed by using an additional high-speed camera that recorded flight kinematics and was synchronized with the PIV. The kinematic analysis showed that during the four wingbeats used to analyze the wake, the bird did not accelerate or decelerate significantly in the streamwise direction.

The wake topology of the starling was characterized using a wake reconstruction based on patterns from the instantaneous vorticity fields where the measurement plane of the velocity was close to the wing root of the bird. The resolution of the data were insufficient to determine if spanwise vortices were present in the near wake or if the observed vorticity was due to the shear created by the flow over the bird's wings. Thus, any connecting spanwise vortices that exist at this stage of the wake development should be relatively small.

The time variation of the profile drag per unit span from classical aerodynamics over each of the four different wingbeat cycles was approximated using the PIV data. Inherent in the calculation of profile drag is the assumption that $\partial u/\partial t$ is zero everywhere or integrates to zero instantaneously, yet this does not necessarily correspond to the case for flapping flight. As with previous studies, the integration of the velocity profiles over the measurement plane admittedly misses the remaining span of the wake, yet clear trends have been observed. It was found that the profile drag term was almost always positive; however, the bird was not observed to noticeably decelerate. Thus, there should be a compensating force to this classical drag term. It was observed that $\partial u/\partial t$ is generally negative during the downstroke from the current dataset; however, the results also show that during the upstroke $\partial u/\partial t$ is generally positive. The approximation of the unsteady term suggests that unsteady aerodynamics may provide some thrust in the overall streamwise force balance.

The role of the unsteady portion of the flow on the flight efficiency of birds is yet to be determined and still remains an open question. Yet, the current results shed light on the role of the unsteadiness during flight and its impact on drag/thrust. In addition, future studies are required to assess how the spanwise variations of these forces affect the balance between drag and thrust.

**Acknowledgements**

Z.J. Taylor gratefully acknowledges the support of the Tel Aviv University Post-doctoral Fellow Scholarship. R. Gurka and C.G. Guglielmo gratefully acknowledge funding from the NSERC Discovery Grants Program, and the Canada Foundation for Innovation and Ontario Research Fund for construction of the AFAR at the University of Western Ontario.

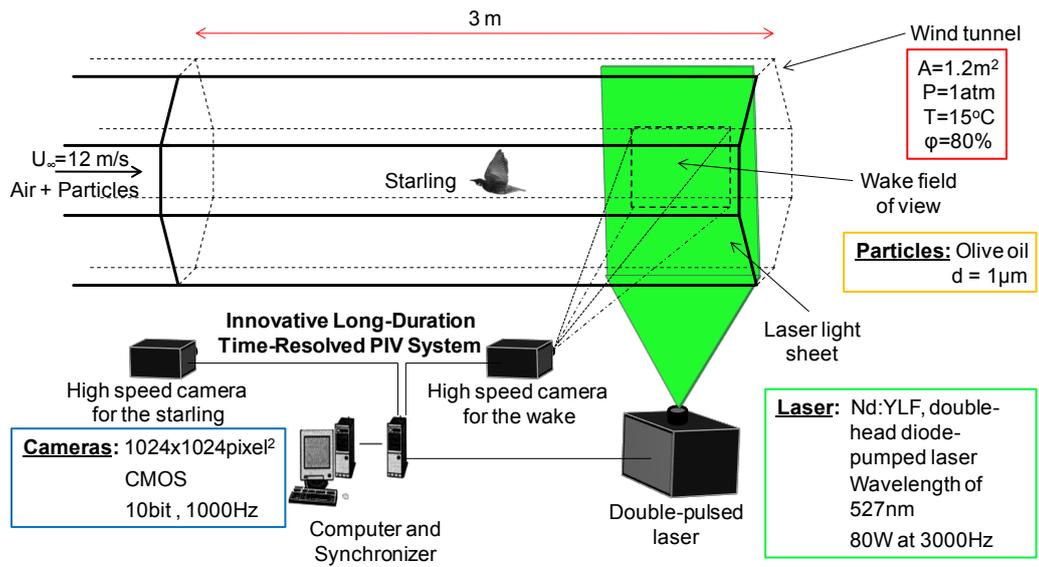

**Figure 1** Illustrative scheme of the experimental setup system.

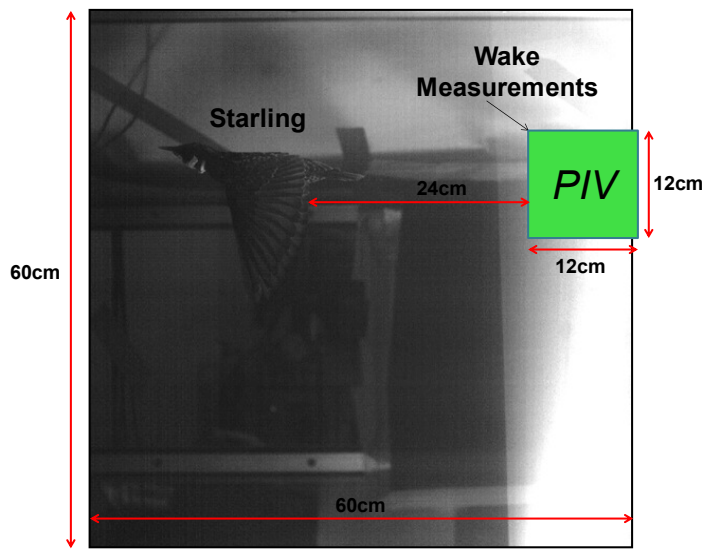

**Figure 2** The large image shows the kinematic camera field of view and the small window marked "PIV" is the PIV camera field of view.

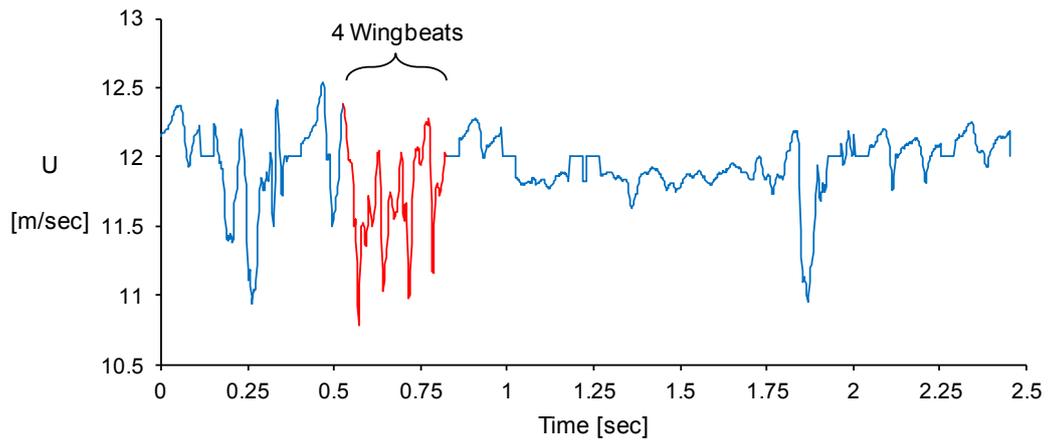

**Figure 3** Variation of the average streamwise velocity in the wake with time.

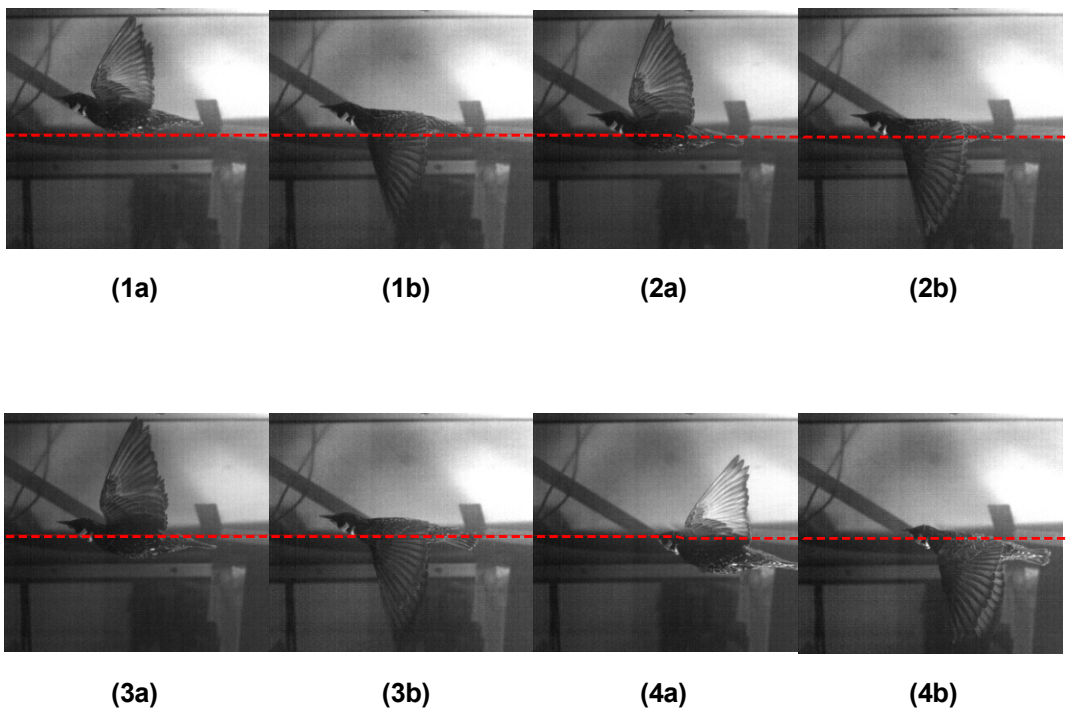

**Figure 4** Different positions of the starling during the four wingbeats. The photographs are labeled so that the number corresponds to the wingbeat number, 'a' marks the beginning of the downstroke, and 'b' marks the beginning of the upstroke.

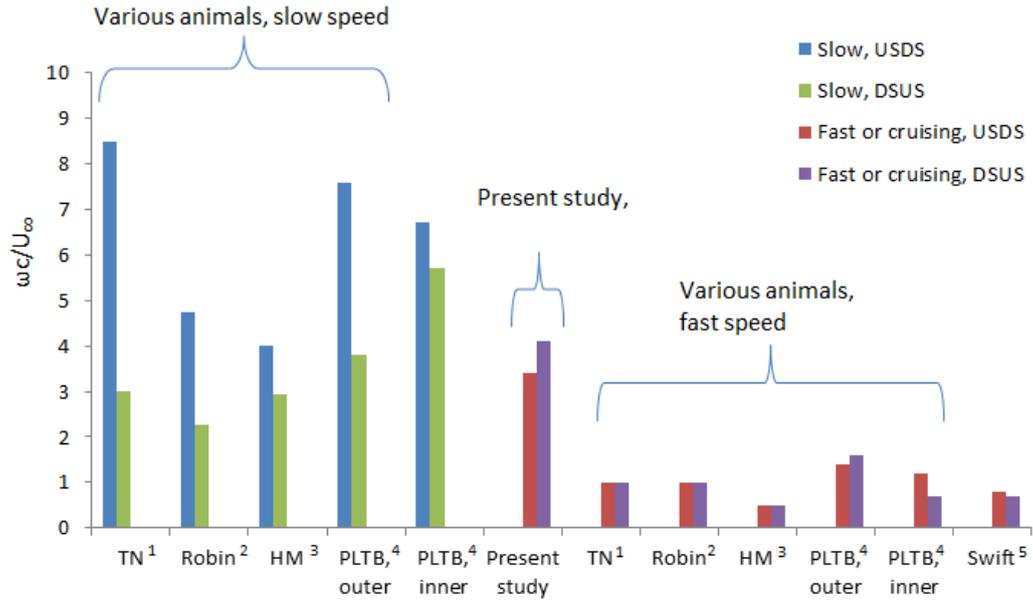

**Figure 5** A comparison of the peak spanwise vorticity measured in the wake of flapping animals from several studies. Superscripts refer to the work of: (1) [18]; (2) [20]; (3) [41]; (4) [1]; and (5) [7]. Abbreviations used in the figure represent the thrush nightingale (TN), house-martin (HM), and Pallas' long-tongued bat (PLTB). Measurements from the Pallas' long-tongued bat come from the inner wing ($z/b_{semi}<0.4$) and outer wing ($z/b_{semi}>0.75$).

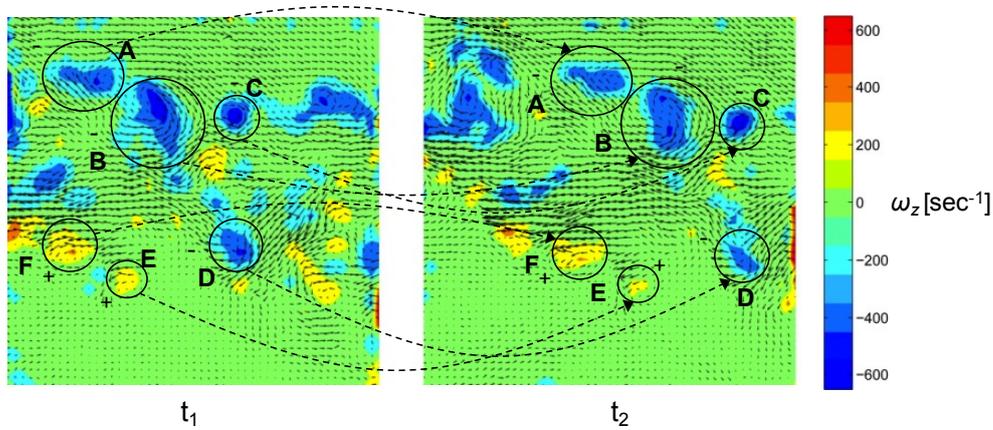

**Figure 6** Two consecutive spanwise vorticity fields ($t_2= t_1+2$ msec). The air flows from left to right and each frame size is 9 x 9 cm$^2$.

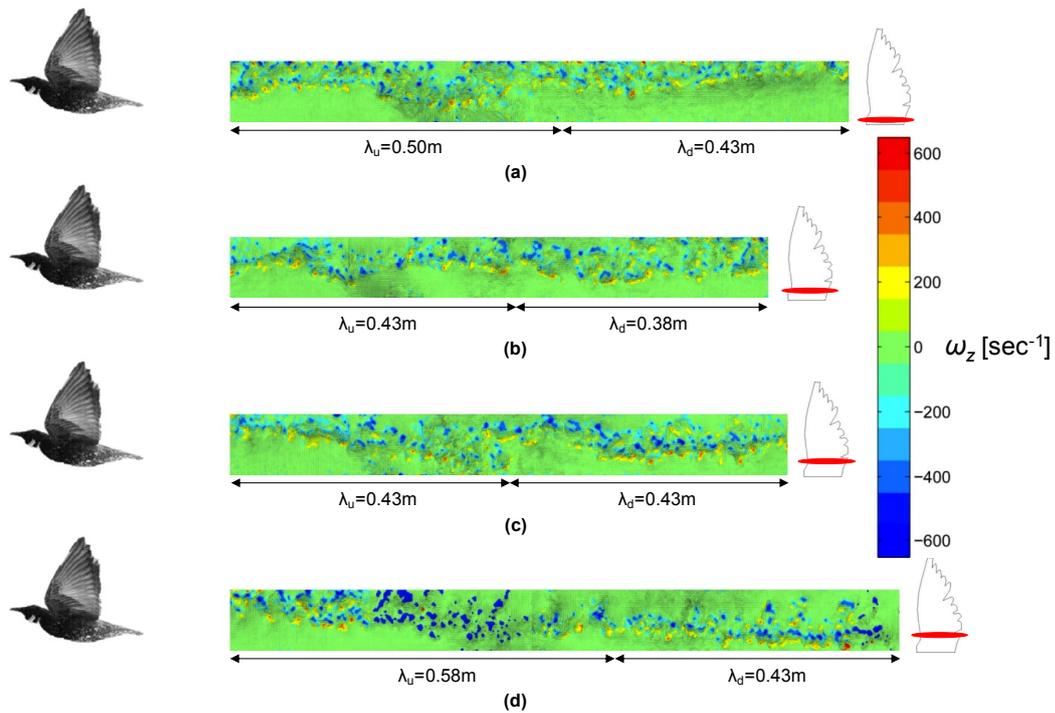

**Figure 7** Reconstruction of the starling's wake consisting of four wingbeats as though the starling flew from right to left. The average spatial flow has been subtracted, thus the vectors displayed are the velocity fluctuations. The contours represent the spanwise vorticity in each wingbeat and the half-wavelengths for the downstroke ($\lambda_d$) and the upstroke ($\lambda_u$) are noted for each wingbeat. Wingbeat numbers go from top to bottom: (a) 1, (b) 2, (c) 3, and (d) 4.

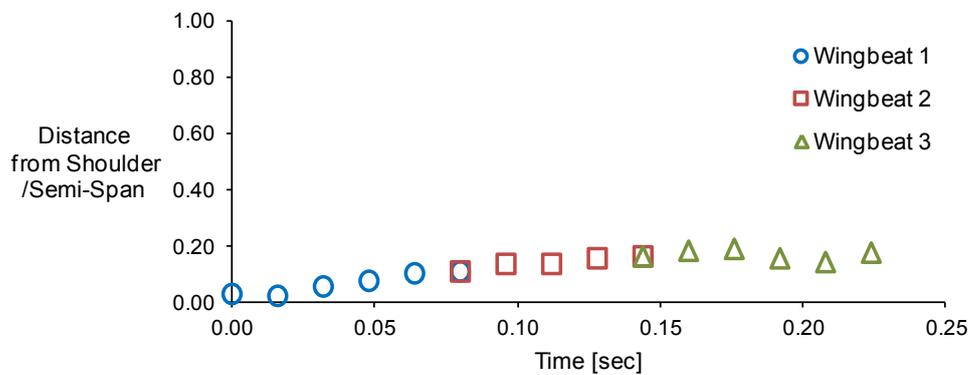

**Figure 8** The location of the laser sheet with respect to the right wing during wingbeats 1, 2, and 3.

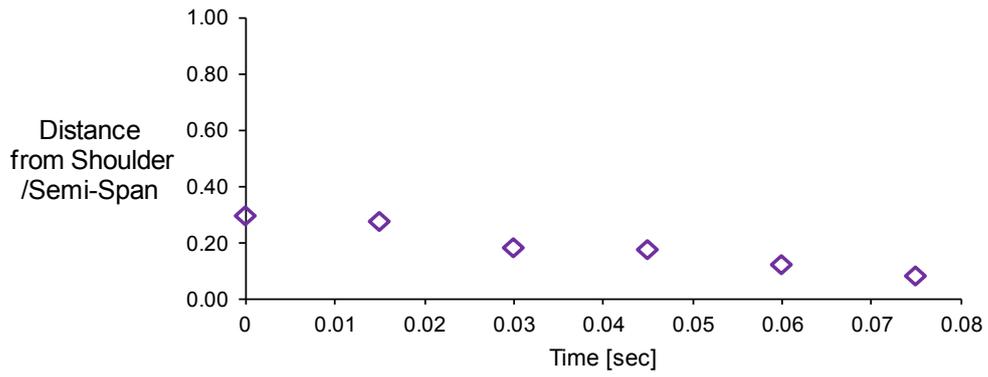

**Figure 9** The location of the laser sheet with respect to the right wing during wingbeat 4.

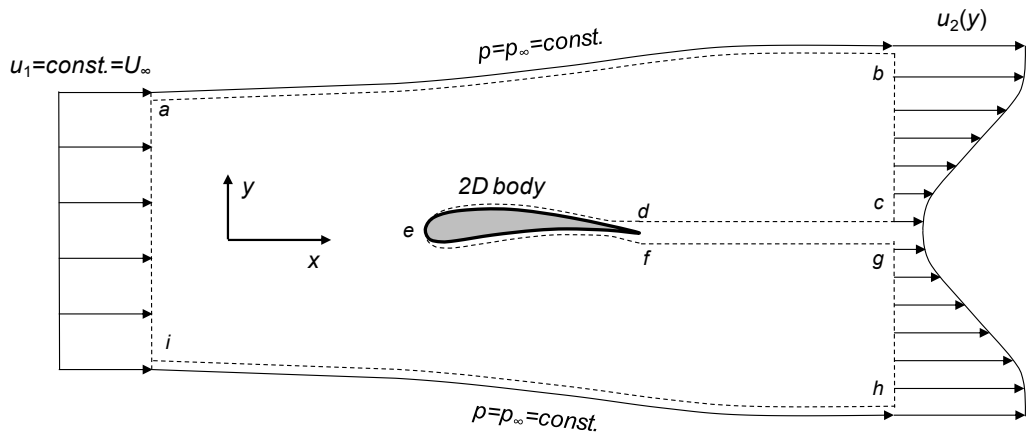

**Figure 10** Control volume around a two-dimensional body in a uniform free stream.

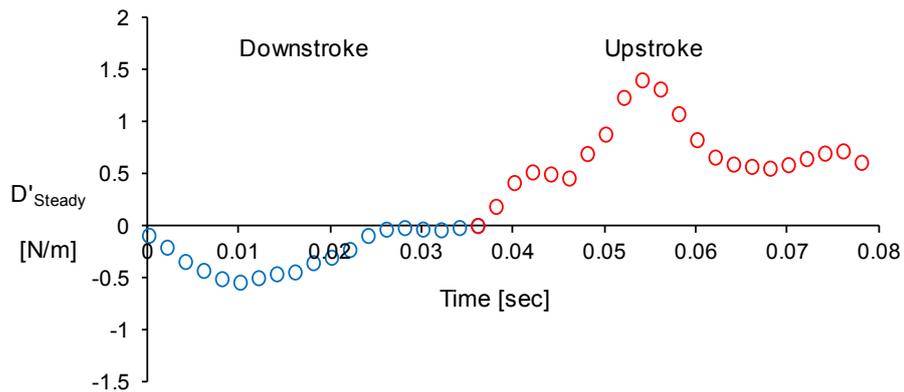

**Fig. 11a** Steady drag per unit span versus time, as computed according to Eq. (13), for wingbeat no. 1.

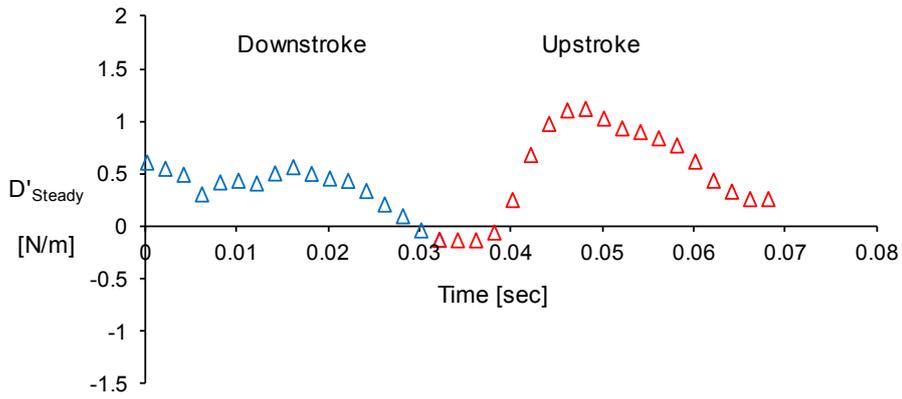

**Fig. 11b** Steady drag per unit span versus time, as computed according to Eq. (13), for wingbeat no. 2.

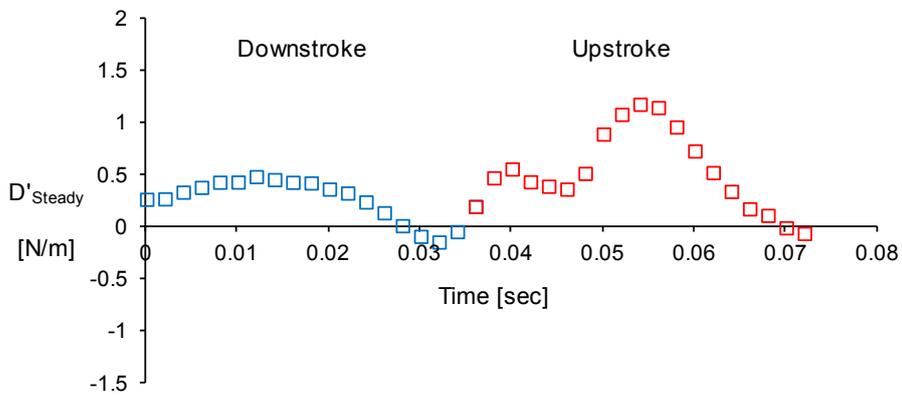

**Fig. 11c** Steady drag per unit span versus time, as computed according to Eq. (13), for wingbeat no. 3.

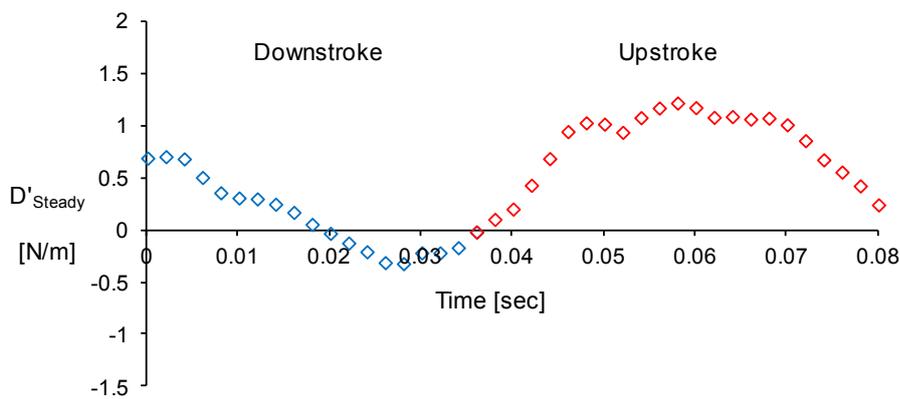

**Fig. 11d** Steady drag per unit span versus time, as computed according to Eq. (13), for wingbeat no. 4.

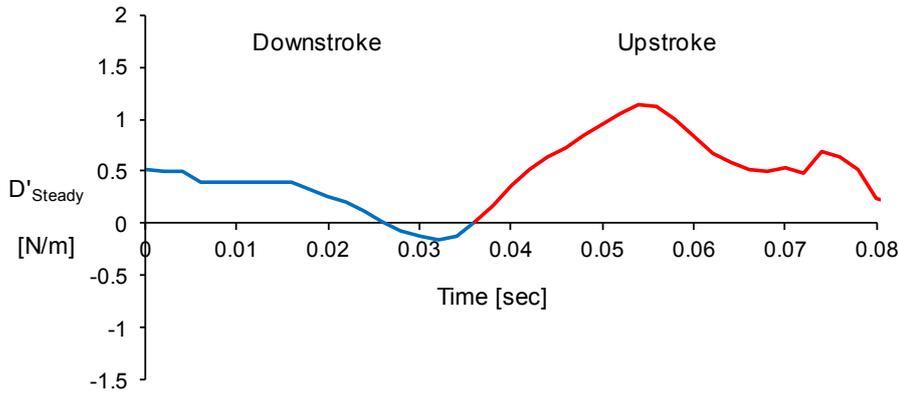

**Fig. 11e** Averaged steady drag per unit span versus time, as computed according to Eq. (13).

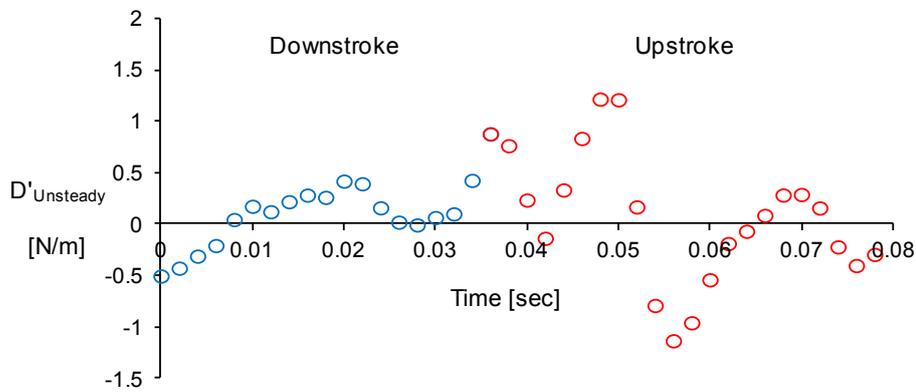

**Fig. 12a** Unsteady drag per unit span versus time, as computed according to Eq. (14), for wingbeat no. 1.

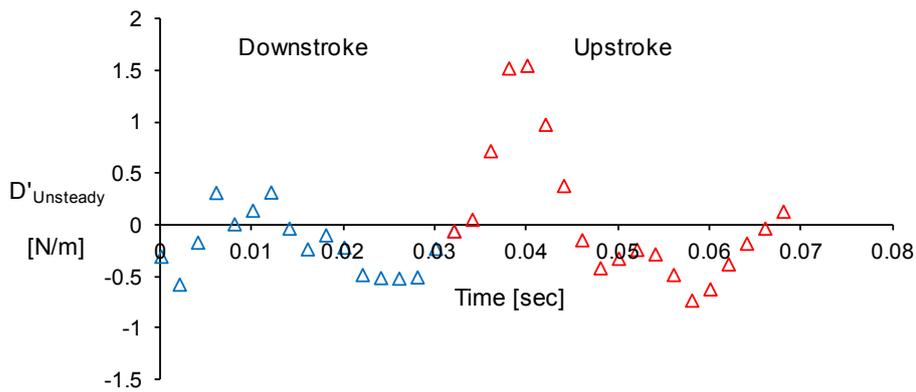

**Fig. 12b** Unsteady drag per unit span versus time, as computed according to Eq. (14), for wingbeat no. 2.

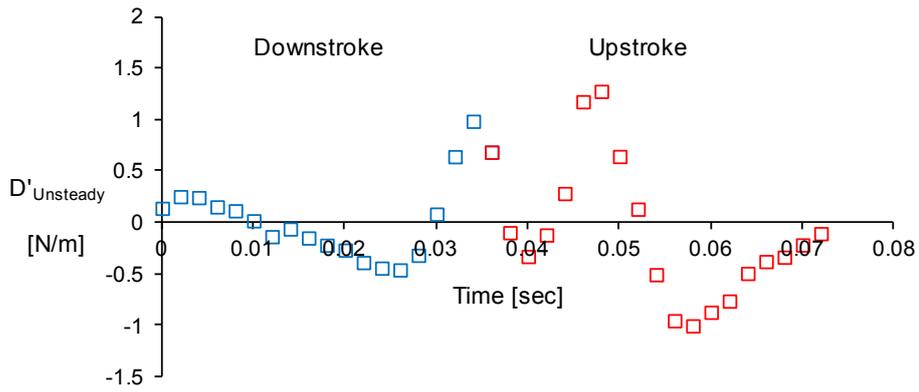

**Fig. 12c** Unsteady drag per unit span versus time, as computed according to Eq. (14), for wingbeat no. 3.

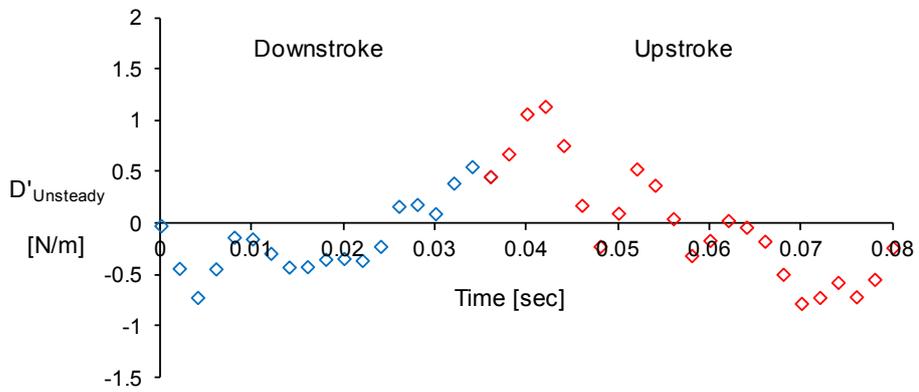

**Fig. 12d** Unsteady drag per unit span versus time, as computed according to Eq. (14), for wingbeat no. 4.

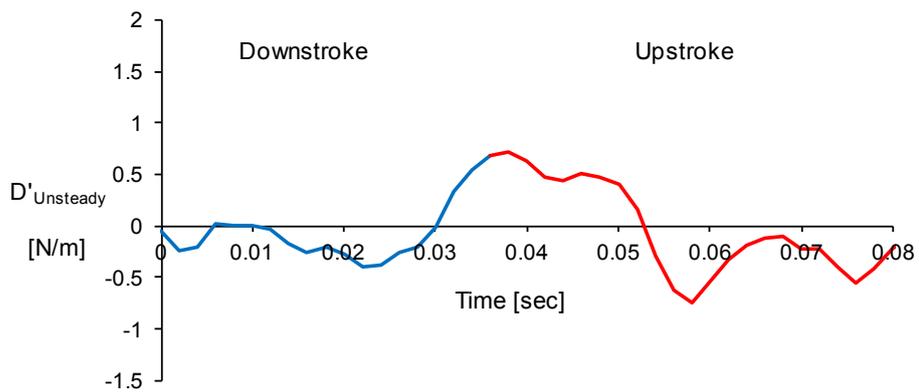

**Fig. 12e** Averaged unsteady drag per unit span versus time, as computed according to Eq. (14).

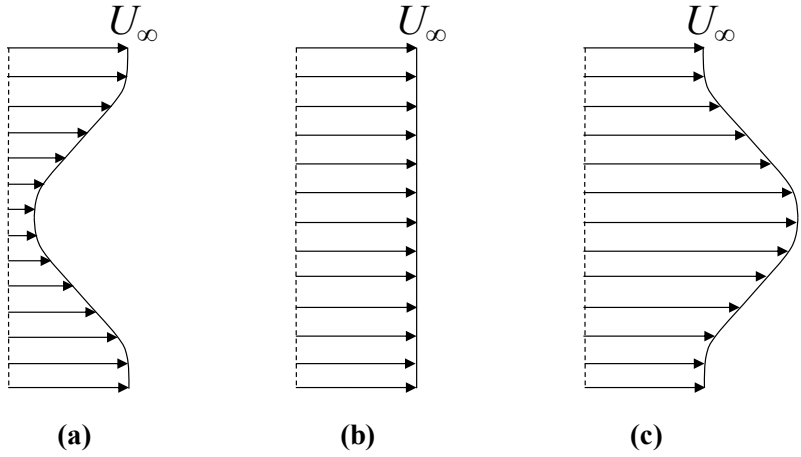

**Figure 13** Schematic examples of a drag wake (a), momentumless wake (b) and a jet wake (c).

**Table 1** The wingbeat frequency, Strouhal number and the wingtip angle of attack for the four wingbeat cycles

| Parameter | Wingbeat | | | | Average* |
|---|---|---|---|---|---|
| | 1 | 2 | 3 | 4 | |
| $f\ [Hz]$ | 12.8 | 14.7 | 13.9 | 11.9 | 13.3 |
| St | 0.27 | 0.34 | 0.33 | 0.24 | 0.30 |
| $\varphi\ [^o]$ | 29 | 34 | 33 | 26 | 31 |

\* The average value of each parameter for the all wingbeats